
\input phyzzx
\nonstopmode
\sequentialequations
\twelvepoint
\nopubblock
\tolerance=5000
\overfullrule=0pt

\REF\bekenstein{J.~D.~Bekenstein \journal Phys.Rev. & D7 (73) 2333}

\REF\hawkinga{S.~W.~Hawking \journal Commun.~Math.~Phys.~ & 43 (75) 199}

\REF\thorne{K.~Thorne, R.~Price, D.~Macdonald
{\it Black Holes: The Membrane Paradigm}, (Yale, 1986).}

\REF\hawkingb{S.~W.~Hawking \journal Phys.Rev. & D14 (76) 2460}

\REF\gibbons{G.~W.~Gibbons and K.~Maeda \journal Nucl. Phys.
& B298 (88) 741}

\REF\garfinkle{D.~Garfinkle, G.~T.~Horowitz and A.~Strominger
\journal Phys.Rev. & D43 (91) 3140}

\REF\limitations{J.~Preskill, P.~Schwarz, A.~Shapere, S.~Trivedi and F.~Wilczek
\journal Mod.Phys.Lett. & A6 (91) 2353}

\REF\hawkingc{J.~B.~Hartle and S.~W.~Hawking \journal Phys.Rev &D13 (76) 2188}

\REF\chand{S.~Chandrasekhar, 1983, {\it The Mathematical Theory
of Black Holes},
(Clarendon Press, Oxford) }

\REF\gilba{G.~Gilbert
{\it On the Perturbations of String-Theoretic Black Holes},
University of Maryland preprint UMDEPP 92-035}

\REF\gilbb{G.~Gilbert {\it The Instability of String-Theoretic Black Holes},
University of Maryland preprint UMDEPP 92-110}

\REF\xanth{B.~C.~Xanthopoulos \journal Proc.~R.~Soc.~Lond.~ &A378 (81) 61-88}

\REF\teit{C.~Teitelboim and F.~Wilczek, {\it Black Hole Atmosphere
and Thermally Induced False Vacuum Decay}, in preparation.}

\REF\rubakov{V.~A.~Rubakov  \journal Nucl.Phys. &B203 (82) 311
\journal Nucl.Phys. &B237 (84) 329 }

\REF\callan{C.~G.~Callan,~Jr. \journal Phys.Rev. & D25 (82) 2141
\journal Phys.Rev. & D26 (82) 2058}

\REF\goldstone{J.~Goldstone, F.~Wilczek  \journal Phys. Rev. Lett. &47 (81)
986}

\REF\callwitt{ C.~G.~Callan,~Jr. and E.~Witten
\journal Nucl.Phys. & B239 (84) 161}

\REF\birrell{N.~D.~Birrell and P.~C.~W.~Davies, 1982, {\it Quantum Fields in
Curved Space} (Cambridge University Press, Cambridge)}

\FIG\potfig{The effective potentials in units $M^{-2}$ versus tortoise $r^*$
in units of $M$ for $l=2$ partial waves, $a=0$, $1$ and $2$ and various
values of the black
hole charge measured in units of the extremal charge $M \sqrt{1+a^2}$,
where M is the black hole mass. For each
set of parameters there are one spectator,
two axial and three polar potentials,
which we display on different plots.
Figure~\potfig a ($a=0$) shows the classical Reissner-Nordstr\"om black
hole which is representative for dilaton black holes with $0\le a <1$.
For $a=1$ (Fig.~\potfig b) the broadening of the potential signals the
emergence
of a mass gap as the black hole becomes extremally charged. $a=2$ (Fig.~\potfig
c) shows the potentials to increase without bound in height while decreasing in
width as the extremal limit is approached, which is
characteristic  for dilaton black holes with $a>1$. Note the changes in scale
for the different values of $a$.}

\FIG\absfig{A typical plot for the dependence of the absorption coefficient
on the frequency of incident waves measured in units of $M^{-1}$.
The parameters
chosen are $a=2$, angular momentum $l=2$ and charge $Q=0$, $0.8$ and $0.99$
measured in units of the extremal
charge $M \sqrt{1+a^2}$ and the plot is for scattering of spectator fields off
a dilaton black hole background.}

\FIG\specfig{A semilogarithmic  plot (base 10) of the  probability of emission
versus frequency measured in units of $M^{-1}$ of a spectator scalar field
in the $l=2$ mode for  $a=0$, $1$ and $2$ and various values of the black
hole charge measured in units of the extremal charge $M \sqrt{1+a^2}$,
where M is the black hole mass. In all cases the spectrum is seen to be
dramatically distorted in the low frequency regime due to grey-body factors.
While the plot for $a=0$, which is the classic Reissner-Nordstr\"om
case representative for dilaton black holes with $0\le a<1$, merely shows a
general decrease in emission  when the black hole becomes extremally charged
--- in accordance with the decreasing temperature ---,
for $a=1$ the development of a mass gap at frequency $5/(4 M)$ predicted from
the widening of the  potential at a value of $3/(2 M)$ is apparent.
For $a=2$ (last plot), the peak of the spectrum reaches after an
initial decrease a constant level, which however shifts to higher frequencies,
as extremality is approached. As discussed in the text, the spectrum must
necessarily be inappropriate for frequencies larger than the black hole mass,
so that the emission turns off when the black hole is extremally charged.}

\def\sm{\scriptstyle}

\line{\hfill IASSNS-HEP-91/71}
\line{\hfill December 1991}
\titlepage
\title{Black Holes as Elementary Particles}
\vskip.2cm
\author{Christoph F.~E.~Holzhey}
\vskip.2cm
\centerline{{\it Joseph Henry Laboratories}}
\centerline{{\it Princeton University}}
\centerline{{\it Princeton, N.J. 08544}}
\vskip.2cm
\author{Frank Wilczek\foot{Research supported in part by DOE grant
DE-FG02-90ER40542}}
\vskip.2cm
\centerline{{\it School of Natural Sciences}}
\centerline{{\it Institute for Advanced Study}}
\centerline{{\it Olden Lane}}
\centerline{{\it Princeton, N.J. 08540}}
\endpage

\abstract{It is argued that the qualitative features of black
holes, regarded as quantum mechanical objects, depend both on the
parameters of the hole and on the microscopic theory in which it
is embedded.  A thermal description is inadequate for extremal holes.
In particular,
extreme holes of the charged dilaton family can have zero
entropy but non-zero, and even
(for $a>1$) formally infinite, temperature.
The existence
of a tendency to radiate at the extreme, which threatens
to overthrow any attempt to identify the entropy as available
internal states and also
to expose
a naked singularity, is at first sight quite disturbing.  However
by analyzing the perturbations around the extreme holes we
show that these holes are protected by mass gaps, or alternatively
potential barriers, which remove them from thermal contact with
the external world.  We suggest that the behavior of
these extreme dilaton
black holes, which from the point of view of traditional black hole
theory seems quite bizarre, can reasonably be interpreted as the
holes doing their
best to behave like normal elementary particles.  The $a<1$ holes
behave qualitatively as extended objects.}

\endpage

\chapter{Introduction and Summary}


Is there a fundamental distinction between black holes and
elementary particles?
The use of concepts like entropy [\bekenstein ], temperature [\hawkinga ],
and dissipative
response [\thorne ] in the description of black hole interactions makes these
objects seem very different from elementary particles.  This has
helped inspire
some suspicion that
the description of the holes may require
a departure from the fundamental principles of quantum mechanics [\hawkingb ].
However a more conservative attitude is certainly not precluded.
In the bulk of this paper, we shall analyze a particular class of
black hole solutions
(extremal dilaton black holes [\gibbons , \garfinkle ])
in some detail, and argue that some of these do in
fact appear to behave very much as elementary particles.  First,
though, let us briefly sketch a radically
conservative interpretation
of black hole physics
in general -- conservative, in the sense that it attempts to
avoid the conclusion that these bodies have radically new and
paradoxical properties, utterly unlike anything we have seen elsewhere
in physics.

The most notable
tangible qualitative physical properties associated with a
black hole's
entropy, temperature, and dissipative response are probably:

\pointbegin
Any projectile impinging within the geometrical radius is almost
certainly absorbed.

\point
The energy thus deposited is taken up by the hole and re-emitted
only after a long time delay.

\point
The form in which the energy is emitted has very little to do
with the details of how it was deposited.  Indeed,
the emission may be
regarded as thermal.

\par

There are other instances in physics of entities with properties
quite similar to those of black holes in these respects.  The liquid
droplet model of atomic
nuclei springs to mind.  Indeed any macroscopic body with many
feebly interacting internal
degrees of freedom, so large that the probability
of a particle traversing the body without being slowed and captured
is small,
might
be expected to exhibit these characteristic behaviors.
Such a body will have many quasi-stable low-lying states, corresponding
to excitations of the internal degrees of freedom which only slowly
dissipate.
Almost regardless of its initial form,
energy upon being absorbed
will be distributed among the low-lying states in a
statistical fashion (thermalized).  This process
of distribution will take a long time
relative to the time an undeflected particle would take to pass through the
body.  Finally, the energy will escape at the surface in a form which has
very little to do with how it was deposited: memory of the initial
excitation
has long since been obliterated.   Thus, each of the main qualitative
features of black hole interactions mentioned above is reproduced by
this general, unmysterious class of physical objects.

There are a couple of obvious
differences between black holes and
liquid drop nuclei, however,
which should prevent one from
accepting the analogy too easily.  First there is the
fact that the gravitational interaction is universal, unlike the
strong interaction.  Thus conceivably there is something more
fundamental about the thermalization process in the case of black
holes, leading to irredeemable loss of information and breakdown
of quantum mechanics (loss of unitary time evolution).  Second
there is the related fact that in the case of the compound
nucleus we have a reasonably good understanding
of how to derive the
semi-phenomenological picture sketched above
as an approximation
to a
more fundamental
microscopic theory which is
manifestly consistent with all general principles
of physics, while in the black hole case there is no comparable
understanding.  The closest approach to such understanding
at present is
probably the ``membrane paradigm'' of black hole physics [\thorne ], which
accurately summarizes many aspects of the low-energy interaction
of black holes in terms of a dynamical surface theory coupled to
the external world.  It is plausible that quantization of a theory
of this kind could give a more satisfactory microscopic understanding
of the internal states of black holes, as states of the effective
surface theory.
(In this regard, it is most suggestive that the
entropy is, to a first approximation, proportional to the surface area
-- that is, an extensive quantity in the effective theory.)
The many feebly interacting degrees of
freedom are the approximate normal modes of this surface theory;
thus, they are collective oscillations of the gravitational
field.
Though much work will be required to substantiate this picture, we feel
there is no evident barrier to taking the proposed analogy quite
literally.

{}From this point of view, the Schwarzschild black holes are
very complicated mixtures of
collective excitations of the gravitational
field.  Their finite temperature indicates that they are far from
the ground state, and their large entropy indicates that they
are embedded in a dense quasi-continuum of states -- in particular,
that
there are many ($\sim e^S = \exp {4\pi GM^2 \over \hbar}$) states within
the typical thermal energy interval
$T = \hbar /(8\pi GM)$ sampled by the
hole.  It therefore seems reasonable to expect that a thermal
description of the hole should be quite appropriate, and that
deviations from it, in particular those
due to quantum effects, should
be small and very difficult to disentangle in practice.
(Significant correlations between emission and absorption, or
between successive emissions, should exist in the rare cases when
such events are separated by small time intervals, of the order of
at most the transit time.)

The situation is quite different for near-extremal black holes, \ie\
holes with nearly vanishing temperature or entropy.  As has recently
been emphasized, and shall be reviewed below,
the thermal description of such holes signals
its own breakdown [\limitations ].  Emission of a single quantum changes the
formal value of the temperature drastically.  More fundamentally,
the total
specific heat of the hole ({\it not\/} the specific heat per unit
volume) becomes small as the hole approaches extremality, which indicates
that the number of distinct states available to the hole in its thermal
energy interval becomes small, so that the foundation for
a statistical description is removed.
It is in this domain that the quantum theory of black holes should
come into its own, and the question whether black holes behave like
more familiar quantum objects or even elementary particles acquires
a sharp edge.


The classic extremal black holes of the Kerr-Newman family
have the character that they have finite entropy at zero temperature.
According to the views outlined above
(\ie\ the conventional view of entropy as it appears elsewhere in
physics), this entropy should be
viewed as an indication of massive unresolved degeneracy for these
holes.  This degeneracy is unusual, but not completely bizarre from
a more microscopic point of view.
After all the hole is capable of absorbing
arbitrarily small amounts of energy
(soft photons or gravitons); thus there are clearly many
low-lying states, and in an approximate calculation (\ie\ the
classical evaluation of black hole mass, which makes no reference
to the state of the 2+1 dimensional membrane theory) these may
appear exactly degenerate.
The degeneracy at the classical level, even if
it can be rationalized in this way, at a minimum makes the
black holes assume more the character of complex extended objects
rather than elementary particles.

Recently it has been discovered that the character of
near-extremal black holes can be affected drastically by the
field content of the world [\gibbons , \garfinkle ].
In particular, inclusion of an
additional scalar degree of freedom (the dilaton), whose existence
has been considered on various ground and is suggested by
superstring theory, with suitable simple couplings has been found
to affect the properties of extremal charged
(Reissner-Nordstr\"om) black holes drastically.
As we shall review
shortly, the relevant coupling of the dilaton is naturally
parametrized by a numerical parameter $a$.  The classic
Reissner-Nordstr\"om solution is unmodified for $a=0$.
For $0<a<1$ the extremal black hole has
zero entropy and zero temperature;
for $a=1$ it has zero entropy and finite temperature;
for $1<a$ it has zero entropy and formally infinite temperature.

In a previous paper [\limitations ] a tentative interpretation of these results
 was
put forward, that is
entirely consistent with the general attitude described above.
For $0<a<1$ the interpretation is straightforward:  zero entropy
at zero temperature indicates that there is a non-degenerate
ground state, which naturally does not radiate.  For $1\leq a$
the situation is more challenging.  It was suggested that
it could be understood only
if the black hole has a {\it mass gap\/} -- that is, if
there are no arbitrarily low-energy excitations around the solution.
For the entropy measures the
density of available states, and with a non-zero temperature it
becomes possible to sample states at any energies of order
the temperature separation.
Furthermore if the formal temperature
is non-zero while the entropy vanishes, we must anticipate that
the mass gap is at least of order the temperature.
If the formal temperature approaches infinity,
as it does for the $1<a$ holes as they approach
extremality, then the mass gap must likewise
approach infinity.

The main burden of the present paper is to demonstrate that these
mass gaps do indeed exist.  We shall perform a complete analysis of
linear perturbations around the charged dilaton
black hole solutions, to demonstrate this
feature.

In plain English, the mass gap means that the black holes have
become universally {\it repulsive\/} rather than attractive for
low energy perturbations.
Thus when probed by scattering of low energy
projectiles, they will appear as elementary particles.  There will
be no time delay, and the outgoing radiation will be strongly,
obviously, and
uniquely correlated with the incoming radiation.
The $a>1$ black holes will respond this way to arbitrarily
high energy probes, at least semi-classically.  The infinite mass
gap also serves to turn off the Hawking radiation, despite the
finite temperature, due to vanishing grey-body factors.





The contents of the remainder of this paper are as follows.  In
Sections 2 and 3 we briefly review the mathematical form of the
dilaton black hole solutions and their (formal) thermodynamic
interpretation, respectively.  In Section 4 we set up the
general perturbation problem for the fields intrinsic to
the solutions, and show how it may be reduced to tractable
form.
That this
is practical at all is a result of some remarkable and unexpected
simplifications.  (In passing, we also clarify and explain some
``miraculous'' aspects of the theory of perturbations about classic
Reissner-Nordstr\"om holes.)
In Sections 5 and 6 we discuss the
physical results which can be inferred from the perturbation
calculation, including quantitative results for grey-body factors
and the mass gaps. The problem of a spectator scalar field is used
to illustrate many of the qualitative results in a much simpler
context.
In Section 7 we briefly note
the anomalous behavior of the Callan-Rubakov mode for scattering of
minimally electrically charged chiral fermions off a magnetically
charged hole, which apparently uniquely does {\it not\/} feel the
repulsive barrier surrounding $a>1$ holes.  Section 8 contains a few
concluding observations.

\chapter{Form of black holes}

We shall be considering
a class of theories involving coupled
gravitational, electromagnetic and
scalar (dilaton) fields.  The action governing
this class of theories
is
$$
S=\int d^4x \sqrt{-g}(R-2(\nabla\Phi)^2+e^{-2a\Phi}F^2)~.
\eqn\forma
$$
We use the metric convention $(+---)$.  $a$ is a dimensionless
parameter, which we may assume to be non-negative.  For $a=0$
\forma\ becomes the standard Einstein-Maxwell action with an extra
free scalar field.  In this case a
solution of the coupled Einstein-Maxwell
equations becomes a solution of the new theory, taking $\Phi ~=~ {\rm
const.}$  The case $a=1$ is suggested by superstring theory, treated
to lowest order in the world-sheet and string loop expansion.

A complete set of equations of motions is readily derived
from this action.
They are
Maxwell's equations
$$
\eqalign{
\partial_\mu(e^{-2a\Phi}\sqrt{-g}F^{\mu\nu})&=0, \cr
F_{\mu\nu,\rho}+F_{\rho\mu,\nu}+F_{\nu\rho,\mu}&=0, \cr}
\eqn\formab
$$
Einstein's equation
$$
R_{\alpha\beta}
=e^{-2a\Phi}(-2F_{\alpha\nu}F_\beta{}^\nu+
{1 \over 2}F^2g_{\alpha\beta})
+2\partial_\alpha\Phi\partial_\beta\Phi ~,
\eqn\formc
$$
and the dilaton equation
$$
{1\over \sqrt{-g}}\partial_\mu(\sqrt{-g}g^{\mu\nu}\partial_\nu\Phi)=
{1\over 2}ae^{-2a\Phi}F^2.
\eqn\formd
$$

Static, spherically symmetric solutions
of these equations have been found, representing charged black holes
[\gibbons , \garfinkle].
The solution for an electrically charged
dilaton black hole has the line element
$$
ds^2 ~=~ \lambda^2 dt^2-\lambda^{-2} dr^2-R^2 d\theta^2-
R^2 \sin^2 \theta d\varphi^2
\eqn\forme
$$
where
$$
\lambda^2 ~=~ \left(1-{r_+ \over r}\right)
        \left(1-{r_- \over r}\right)^{1-a^2 \over 1+a^2}
\eqn\forf
$$
and
$$
R^2~ = ~r^2 \left(1-{r_- \over r}\right)^{2a^2 \over 1+a^2} .
\eqn\formg
$$
In these equations $r_+$ and $r_-$ are the values of the
parameter $r$ at the outer and the inner horizon, respectively,
where the coefficient of $dt$ vanishes and the flow of
global time $t$ has no influence.  $r_+$ and $r_-$ are related to
the mass and charge of the hole according to
$$
2M ~=~ r_+ + \left({1-a^2 \over 1+a^2}\right) r_-
\eqn\formh
$$
and
$$
Q^2 ~= ~{r_- r_+ \over 1+a^2}~.
\eqn\formi
$$
Finally, the electric and dilaton fields are given by
$$
e^{2a\Phi} ~= ~ \left(1-{r_- \over r}\right)^{2a^2 \over 1+a^2}
\eqn\formj
$$
and
$$
F_{tr} ~=~ {e^{2a\Phi}Q \over R^2}  ~.
\eqn\formk
$$
Note that
$R$, not $r$, has the normal meaning of the radial variable,
in the sense that the area of the sphere obtained by varying
$\theta$ and $\phi$ at a fixed $t$ and $r$ (or $R$) is $4\pi R^2$.

There are also magnetically charged solutions of the same general form.
They are obtained from the electric solution by making the changes
$$
\eqalign{
F '_{\mu\nu}
   ~&=~ e^{-2a\Phi}
  {1\over 2}{\epsilon_{\mu\nu}}^{\lambda\rho}F_{\lambda\rho}\cr
\phi ' ~&= ~ -\phi ~.\cr}
\eqn\forml
$$


Particularly interesting is the structure of horizons,
and the structure of extremal holes.

For $a=0$ there
are both inner and outer horizons, at $r~=~r_{\pm}$.
The geometry is not singular at either of these.  However
for $a>0$ the geometry does become singular at $r_-$, because
$R$ vanishes there.  In fact $r~=~r_-$ is a spacelike surface
of singularity, very similar to the singularity in the classic
Schwarzschild metric.


This difference gives rise to a big difference in the structure of
the extremal black holes (defined as those on the verge
of exposing a naked singularity) in the two cases.
In all cases the extremal holes occur when $r_+~=~r_-$, which occurs
for
$$
M^2 ~=~ {Q^2 \over (1+a^2)} ~~ {\rm (extremal)}~.
\eqn\formm
$$
For $a=0$ the physical radius $R$ ($=r$) of the horizon is finite,
in fact equal to $M$, at this point.  Further increase of $Q$
at fixed $M$ induces
$r_{\pm}$ to wander off into the complex plane, exposing the
pre-existing singularity at $r=0$.  For $a>0$ the physical radius
of the horizon vanishes for the extremal hole, and the geometry is
singular there. Further increase of $Q$ at fixed
$M$ leads to
the singular ``inner'' horizon $r_-$ extending outside the
ordinary horizon $r_+$, which however both remain real (forever
if $a \geq 1$; until $Q^2 = M^2/(1-a^2)$ if $a < 1$).


The distance to the horizon is infinite for the classic
extremal Reissner-Nordstr\"om hole, although it can be traversed
in finite proper time.  For $a>0$ the distance remains finite,
that is
$$
\int_{r_+}~{dr\over \lambda} ~<~ \infty~.
\eqn\formn
$$
For $a>1$ the tortoise co-ordinate, which is the appropriate
co-ordinate for analyzing the wave equation for external fields
interacting with the hole, has a {\it finite range}.  That
is,
$$
\int_{r_+}~dr^* ~\equiv ~
\int_{r_+}~{dr\over \lambda^2} ~\sim~
\int_{r_+}~{dr\over (r - r_+)^{2\over (1+a^2)}} ~<~ \infty~~ (a>1)~.
\eqn\formo
$$

The fact that the tortoise co-ordinate does not extend to
$-\infty$ for extremal holes when $a>1$ has an important
implication.  The
proof of the usual classical no-hair theorems rely crucially
on the fact that for ordinary black holes the tortoise co-ordinate,
which is the co-ordinate that governs the wave equation for
perturbations, extends to $-\infty$ at the horizon.  It is this
fact which prevents the existence of
finite ``Yukawa tails'' starting at the horizon.  Such arguments
evidently do not apply to the extremal holes under discussion,
and other considerations are required
to address the
question of whether classical hair exists
for these holes.
We shall see below that there are infinite
potential barriers surrounding the hole, which forbid the existence
of at least some forms of hair; but the answer in principle
depends on non-universal aspects of the coupling of the hole to
external fields.  In any case, the fact that the
horizon is in this profound sense only a finite distance away
is certainly suggestive of the elementary particle interpretation
of these holes we are developing.

\chapter{Thermal interpretation}

\section{Temperature and entropy}

The quickest way to find the temperature of a static
black hole
is to consider its behavior for imaginary values of the
time [\hawkingc ].  As is well known, for the classic
black hole the horizon then represents a conical
singularity of the resulting solution of the Euclideanized
Einstein equations, which can be removed if the imaginary time
variable is taken to be periodic with just the right period.
The nonsingular configuration can be regarded as a contribution
to the partition function for temperature equal to the inverse
of the imaginary time period.  Alternatively, the periodicity
shows that fields in
the black hole geometry can be in thermal equilibrium
with a heat bath at that temperature at spatial infinity.
The same procedure works, without modification, for the
dilaton holes.  For the temperature, one finds
$$
T
={1\over 4 \pi r_+}\left({r_+ - r_-\over r_+}\right)^{{1-a^2\over 1+a^2}}~.
\eqn\therma
$$
The entropy may then be inferred from the second law of thermodynamics,
or alternatively, following the argument sketched above,
computed from the interpretation of the black hole
as a saddle-point contribution to the partition function.
Both methods give the same result, to wit
$$
S=\pi {r_+}^2\left({r_+ - r_-\over r_+}\right)^{{2a^2\over 1+a^2}}
= {1\over 4} A~.
\eqn\thermb
$$
where $A$ is the area of the event horizon.

Evidently the extremal holes, for which $r_+ = r_-$, will have
finite entropy in case $a=0$, but zero entropy otherwise.
The temperature of the extremal holes is zero for $a<1$, finite
and equal to $1/(8\pi M)$
(the same value as for a Schwarzschild hole) for $a=1$,
and infinite for $a>1$.

\section{Grey-body factors}

Assuming for the moment that a thermal description is appropriate,
so that each black hole can be in equilibrium with a thermal bath
at its characteristic temperature, there still remains the question
of how well it couples thermally to the external world.  In other words,
to calculate how fast it exchanges energy in a given mode with the
external world, and thereby to a first approximation how fast it
will radiate that mode into empty space, one must compute the
appropriate grey-body factor.  This is essentially the
square of the absorption amplitude for the mode.  More precisely,
rate of radiation is given by
$$
{dM\over dt} ~=~ - \Sigma_{{\rm modes}~\alpha}
   \int ~{d\sigma \over 2\pi} ~ B_{\alpha, \sigma}
   ~(1 - | R_\alpha (\sigma )|^2 ) \hbar \sigma~,
\eqn\thermc
$$
where $\sigma$ is the frequency of the mode in global time $t$, and
$B$ is the Boltzmann occupancy factor.  (Thus
$B =  1/(\exp (\hbar \sigma / T )\pm 1) $ for uncharged fermions and
bosons, respectively.  For charged modes, or for modes carrying angular
momentum in the background of a Kerr hole,
the Boltzmann factor would of course be modified by the
appearance of appropriate
chemical potentials.)   $R_\alpha$ is the reflection amplitude, that is
the amplitude of the wave reflected back to $r_* = -\infty $
relative to the amplitude of the wave that propagates to $r_* = +\infty$,
subject to the requirement
that there is no amplitude for waves travelling in from
infinity.

\section{Breakdown of thermal description}

There are very significant circumstances, in which even the
approximate
description of a black hole
as a thermal body is not physically consistent [\limitations ].
If the emission of a single quantum of the ``typical''
radiation product, which of course changes the mass of the
black hole, also changes the formal value of its temperature
by an amount that is large relative to the temperature itself,
then the notion of the Boltzmann factor for its emission, and
concretely equation \thermc\ , becomes ambiguous.  One does not know
whether to use the temperature before emission, the temperature
after, or something in between.  This would seem to be a rather
clear signal that the effect of
back-reaction on the black hole geometry cannot
be neglected, even approximately, in describing radiation from
the hole.

As was discussed extensively in [\limitations ], the circumstance just
mentioned
occurs, and the thermal description of the hole breaks down,
whenever
$$
| ~\left( {\partial T\over \partial M} \right)_{Q,J}~| ~\geq~ 1 ~.
\eqn\thermf
$$
An alternative form of this condition is
$$
T\left( {\partial S\over \partial T } \right)_{Q,J} ~\leq~ 1~.
\eqn\thermg
$$
The alternative form has a profound physical interpretation.
It says that the thermal description must break down when
the {\it available entropy} of the black hole, \ie\ the number of
states available to it within its thermal energy interval, is
small.  Indeed, if that happens clearly a statistical description
of the physics, which then involves only a few degrees of
freedom, becomes inappropriate.

Upon evaluating the left-hand side of \thermf\
for near-extremal holes we find in the $a=0$ case
$$
| ~\left( {\partial T\over \partial M} \right)_{Q,J}~|
{}~\rightarrow ~ {1\over 2\pi M^2} (1~-~ {Q^2\over M^2}) ^{-{1\over 2}}
{}~\rightarrow ~ \infty ~,
\eqn\thermh
$$
as $(1 - {Q^2\over M^2}) \rightarrow 0$.  And for
$a\neq 0$ we find from \formh\ and \formi\ that
$$
r_+ - r_- ~\rightarrow ~ {\rm const.} (M^2 - {Q^2\over 1 + a^2} )
\eqn\thermi
$$
as $f\equiv 1 - {Q^2\over M^2 (1 + a^2 )} \rightarrow 0$, where the constant
is non-singular in this limit.  From this we find that
$$
T ~\sim~ f^{1 -a^2 \over 1+a^2}
\eqn\thermj
$$
and
that $({\partial T\over \partial M })$ diverges, except in the
special case $a=1$.  The sign of this diverging quantity is of
some interest: it indicates that for $a<1$ the black hole has
a positive specific heat (that is, it cools as it radiates)
as it approaches extremality, whereas for $a>1$ the specific heat
is negative.

The physical reason for the breakdown of a thermal description,
as we have said, is the sparsity of nearby states.  Now ordinarily one
would think that a black hole has a great variety of nearby states --
after all, it seems it should be possible to dump arbitrarily
small amounts
of energy into the hole in different partial waves, ... .
How could this expectation fail?
Actually there are two different ways it could fail, which are
realized for $a<1$ and $a\geq 1$ ({\it including}
$a=1$)
extremal holes respectively.

It can fail if the notion of
what is a ``nearby'' state becomes very demanding -- \ie\ if the
thermal interval becomes very small, or in plain English if the
temperature vanishes.  (It might seem that we have proved too much
here, because thermodynamics has had some famous successes in
describing low-temperature physics.  However, as was mentioned in
[\limitations ], it is true as a matter
of principle that the domain of validity of a thermal description,
{\it for a body of fixed size}, shrinks to zero as the temperature
approaches zero.  By passing to arbitrarily large bodies we create
more and more low-energy states, and thermal states at lower and
lower temperatures become meaningful.  This is of course not possible
for a black hole, which is a single object with a definite size.)
This is what happens for $a<1$.

It can also fail even if the measure of ``nearby'' is generous --
if the temperature is finite or even formally divergent -- if
there is a sufficiently large {\it mass gap\/} for the hole.
The existence of such a gap  entails among other things that
it must {\it not} be possible to dump arbitrarily small amounts of
energy into the hole in many ways -- that the black hole effectively
{\it repels} low-energy perturbations!  As we shall see, this is
what happens for $a\geq 1$.

Another aspect of the extremal black holes is crucially related
to the existence of the mass gap.  That is the question whether they
really are endpoints of Hawking radiation, or whether they continue to
radiate.  Of course if
they did continue to radiate, they would expose a naked
singularity.  However the calculations that follow make it very
plausible that the radiation shuts off.  For $a<1$ of course
there is nothing to show, because the temperature vanishes
for the extremal hole.
For $a>1$ the temperature of the extremal hole
is formally infinite, as is the mass gap.  To elucidate the physics,
one must consider how this limit is approached.  We shall
show that as
the hole approaches extremality its grey-body factors kill all
the
radiation below a critical energy, so that the black-body emission
spectrum is extremely distorted.  The critical energy is always
larger than the temperature, and in particular it becomes arbitrarily
large -- larger than the mass of the hole --
as extremality is approached.
Again, therefore, the radiation shuts off.
The case $a=1$ remains enigmatic.

It is important to note that in all cases the
grey-body factor for low energy radiation
is never quite zero for a non-extremal hole,
so that the radiation never
quite turns off.  Thus the extremal hole is in fact approached,
although slowly, as the radiation proceeds.

\chapter{Formulation of perturbation equations}

We turn now
to the formulation of the equations satisfied by small perturbations
of the fields present in the stationary configuration, that is the
metric functions, the electromagnetic field and  the dilaton field.
For concreteness,
we shall first study the case of an electrically charged dilaton
black hole.

Since the background configuration
is stationary and spherically symmetric, the
most general metric that need be considered for first order perturbations
will be nonstationary but may be assumed to be axially symmetric.
According to [\chand ], it can be written in the form:

$$g_{\mu\nu}=\left( \matrix {e^{2\nu} & \omega e^{2\psi} & 0 & 0 \cr
     \omega e^{2\psi} &-e^{2\psi} & q_2e^{2\psi}&q_3e^{2\psi}   \cr
     0 & q_2e^{2\psi} & -e^{2\mu_2} & 0 \cr
     0 & q_3e^{2\psi} & 0 & -e^{2\mu_3}  \cr } \right).
\eqn\metric$$
in the coordinate basis $x^\mu = (t,\varphi,r,\theta)$.
It will prove convenient to
express tensors in terms of a tetrad basis $e_\alpha^{(a)}$,
chosen so that
$e_\alpha^{(a)}e_\beta^{(b)}\eta_{(a)(b)}=g_{\alpha\beta}$ where
$\eta_{\alpha\beta}={\rm diag}(+1,-1,-1,-1)$. The explicit form
of the
tetrads for the above metric is ([\chand ], p. 81):
$$\matrix{
e_\alpha^{(0)}&=\bigl(&e^\nu&,&0&,&0&,&0&\bigr)\cr
e_\alpha^{(1)}&=\bigl(&-\omega e^\psi&,&e^\psi&,&-q_2 e^\psi&,
                &-q_3e^\psi&\bigr)\cr
e_\alpha^{(2)}&=\bigl(&0&,&0&,&e^{\mu_2}&,&0&\bigr)\cr
e_\alpha^{(3)}&=\bigl(&0&,&0&,&0&,&e^{\mu_3}&\bigr)\cr
}\eqn\tetr$$
Henceforth, we will usually refer to tensors in their tetrad basis.
To make this
clear we will use Roman indices for general components and Arabic
numbers for
specific components, while continuing to use
Greek indices and coordinate
names to refer to tensor components in the coordinate basis.
The transformation between the two different frames is achieved by the use of
tetrads, so that we have for example
$F_{\mu\nu}=F_{ab} e_{\mu}^a e_{\nu}^b$.

Given that the background configuration described
in Section 2 consists of a diagonal
metric, a non-zero dilaton field and $F_{02}=Q e^{2a\Phi} R^{-2}$,
the perturbed configuration is parametrized by small changes
$\delta \nu$, $\delta \psi$, $\delta \mu_2$, $\delta \mu_3$,
$\delta F_{02}$ and $\delta \Phi$ of the background fields, as well as
small values of $\omega$, $q_2$, $q_3$ and $F_{ab}$  for
$\sm{(ab)\not= (02)}$.  This yields a total inventory
of fourteen functions.
Fortunately, upon substituting these functions into the equations
of motion and  keeping only terms of
first order in the perturbations,
one observes that the linearized
equations of motion naturally fall into two sets.  One set
relates
$\delta F_{02}$, $F_{03}$, $F_{23}$, $\delta \nu$, $\delta \psi$,
$\delta \mu_2$, $\delta \mu_3$ and $\delta \Phi$, while the other relates
$F_{01}$, $F_{12}$, $F_{13}$, $\omega$, $q_2$ and $q_3$.
The first set of equations describes polar perturbations, that preserve
spherical symmetry, while
the second one describes so-called axial perturbations.
(Because nonzero
$\omega$, $q_2$ and  $q_3$ imply the loss of spherical symmetry, leaving only
axial symmetry.)
This standard terminology is taken over from [\chand ], where
the same separation was
employed for the study of perturbations of classic Reissner-Nordstr\"om
($a=0$) black holes.
The polar equations prove to be much harder to analyze, even for the
classic Reissner-Nordstr\"om holes.
It is noteworthy that dilaton perturbations do not appear
in the axial equations,
which can therefore be treated in the general case
with little more difficulty
than
in the classic case.

\section{The axial equations}

The axial equations are obtained by linearizing the first
Maxwell equation
for $\nu=\varphi$, the second Maxwell equation for
$\sm{(\mu\nu\rho)=(\varphi tr)}$ and $\sm{(\varphi t \theta)}$ and the
Einstein equation for $\sm{(ab)=(12)}$ and $\sm{(13)}$.
There are in fact two more axial equations but they turn
out to be redundant.
The analysis required to derive
the axial equations in their explicit form precisely parallels
the procedure outlined in [\chand ],
so we will
restrict ourselves to a brief
sketch. The partial
differential equations in the variables $t$, $r$
and $\theta$
separate if we
look for solutions with the time dependence $e^{i\sigma t}$ and make the
following {\it ansatz\/} for the radial and angular part
(the angle $\varphi$ does not enter, due to axial symmetry):

$$\eqalign{
\left(q_{2,\theta}-q_{3,r}\right) &={\tilde Q(r) C_{l+2}^{-3/2}\over R^2
 \lambda^2
                                    \sin^3\theta} \cr
F_{01}&={B(r) \over \sin^2\theta} {dC_{l+2}^{-3/2}  \over d\theta} ,\cr}
\eqn\subsa$$
where $C_n^\nu$ denotes the Gegenbauer function, which solves the
equation
$$\left({d\over d\theta}\sin^{2\nu}\theta {d\over d\theta}+n(n+2\nu)
\sin^{2\nu}\theta \right)\quad C_n^\nu(\theta)=0.$$
With this substitution the five first order partial differential equations are
reduced to two coupled ordinary second order differential equations:

$$\eqalign{
e^{2a\Phi}\left(e^{-2a\Phi}\lambda^2 (\lambda RB)_{,r}\right)_{,r}+
\left(\sigma^2 R \lambda^{-1}- \left(\mu^2+2\right)\lambda R^{-1}-
4 Q^2 \lambda e^{2a\Phi} R^{-3}\right)B &\cr
- Qe^{2a\Phi} R^{-4} \tilde Q(r) &=0 \cr
R^4 \left(\lambda^2 R^{-2} \tilde Q(r)_{,r}\right)_{,r}
+\left(\sigma^2 R^2 \lambda^{-2} -\mu^2\right)\tilde Q(r)
-4\lambda Q\mu^2 R B &=0,\cr} \eqn\ax$$
where $\mu^2 \equiv l(l+1)-2$.
The first order derivatives can be eliminated by the change of variables
$$\tilde Q(r)=RH_2 \quad,\quad R\lambda B=-H_1 e^{a\Phi}/2\mu \quad {\rm and}
\quad dr=\lambda^2 dr^*.$$
We recall that the tortoise coordinate $r^*$ tends for non-extremal black
holes to $-\infty$ as the horizon is approached, whereas it behaves like
$r+ 2M\ln r$ for large radii.

Thus we arrive at two one-dimensional wave equations
coupled by
an interaction matrix:
$$
\left({d^2 \over dr_*^2} +\sigma^2\right)
\left(  H_1 \atop H_2 \right) =
\left( {V_1 \atop V_{12}} {V_{12} \atop V_2} \right)
\left(  H_1 \atop H_2 \right)~,
\eqn\axi$$
where
$$\eqalign{
V_1&=a^2 (\Phi_{,r^*})^2-a \Phi_{,r^*r^*}+ (\mu^2+2)\lambda^2  R^{-2}+
 4 Q^2 \lambda^2  e^{2 a\Phi} R^{-4}\cr
V_2&= 2 R^{-2} (R_{,r^*})^2 - R^{-1} R_{,r^*r^*}
+ \mu^2 \lambda^2 R^{-2} \cr
V_{12}&=-2 Q \mu e^{a\Phi} \lambda^2 R^{-3}. \cr}
\eqn\axpot$$
The interaction matrix can in fact be diagonalized
by a coordinate-independent
similarity transformation so that two independent
wave equations are obtained, with potentials $U_1$ and $U_2$ given by
$$U_{1,2}={1\over 2} \left(V_1+V_2\pm \sqrt{(V_1-V_2)^2+4V_{12}^2}\right).$$

These equations are
roughly of the same form as those derived by Gilbert [\gilba ]
in the
special case $a=1$, but the differences  are crucial.
Gilbert found that
the potentials generically have negative portions,
from which he concluded that
charged dilaton black holes are unstable [\gilbb ].
As we shall soon see,
however, the potentials for non-extreme black holes
go to zero both at the horizon and at large radii
and are {\it positive} in the
intermediate region. \footnote*{
The main error in Gilbert's derivation seems to an algebraic slip:
Substitution of his equations
(G55) and (G56) into (G49) leads to (G57) with
${\cal T}_1=-\left(\Phi_{,r^*r^*}+\Phi_{,r^*}{}^2\right)$ instead of (G59).
With this definition of ${\cal T}_1$, (G62-66) agrees with \axi -\axpot\
except for a sign
change in $\Phi$, which seems to indicate that he is effectively
dealing with a magnetically charged hole. }

Since the gravitational waves have spin two,
it should not come as a surprise that the
analytic form of the equations
changes for $l\leq 1$.
Equation \subsa\ shows that $\tilde Q(r)$
must be zero identically for $l<2$ lest the metric functions $q_2$ or $q_3$
diverge at $\sin \theta =0$. The two equations \ax\ then
become inconsistent for $l=0$ unless $B \equiv 0$, whereas for $l=1$
the second equation of \ax\ is identically satisfied, leaving a
wave equation for the electromagnetic mode with the potential $V_1$
having $\mu^2=0$.

\section{The polar equations}

As in the axial case we will choose a subset of all the polar equations,
namely the first Maxwell equation
for $\nu=r,\theta$, the second Maxwell equation for $\sm{(\mu \nu \rho)=
(tr\theta)}$ the Einstein equation for $\sm{(ab)=(02)}$, $\sm{(03)}$,
$\sm{(23)}$, $\sm{(11)}$, $\sm{(22)}$, and the dilaton equation.
The other
polar equations are redundant,
as, in fact, is the $R_{11}$-equation which
however we retain for later convenience.
Following again closely the methods described in [\chand ],
we separate the partial differential equations by ascribing
an $e^{i \sigma t}$ time dependence to all perturbations
$\delta F_{02}$, $F_{03}$, $F_{23}$, $\delta \nu$, $\delta \psi$,
$\delta \mu_2$, $\delta \mu_3$ and $\delta \Phi$
and making the following {\it ansatz\/} for their angular dependence:

$$\eqalign{
\delta \nu&=N(r) P_l(\theta) \qquad
\delta\psi=T(r) P_l(\theta) +2 \mu^{-2} X(r)P_l(\theta)_{,\theta}
 \cot\theta \cr
\delta\mu_2&=L(r)P_l(\theta) \qquad
\delta\mu_3=T(r)P_l(\theta)+2 \mu^{-2} X(r)P_l(\theta)_{,\theta\theta}\cr
\delta F_{02}&=-{R^2 \lambda^2 \over 2 Q} B_{02}(r)P_l(\theta) \qquad
F_{03}={R \lambda \over 2 Q} B_{03}P_l(\theta)_{,\theta} \cr
F_{23}&= {i\sigma R \over 2 Q \lambda} B_{23}P_l(\theta)_{,\theta} \qquad
\delta\Phi=\phi P_l(\theta),}\eqn\psub$$
where $P_l(\theta)$ are Legendre polynomials with
index $l$ which we write in terms of $\mu^2 \equiv l(l+1)-2$.
Substituting these functions into the equations of motion, using
Chandrasekhar's expressions for the $R_{ab}$'s and linearizing in the
perturbations, the following equations are
obtained.  (They are listed in the order they were enumerated
above)\footnote*{It is in fact more convenient to use the $G_{22}$-equation
instead of the $R_{22}$-equation to obtain (4.15).
For the $R_{11}$-equation,
which is a sum
of two terms with different angular dependences, both coefficients
have to separately vanish. Only  one of these two equations is displayed
as (4.14).}.

$$\lambda^2B_{02}-4Q^2e^{2a\Phi}R^{-4}(B_{23}-L-X-a\phi)
+(\mu^2+2)R^{-2}B_{23}=0 \eqn\pola$$
$$B_{03}-e^{2a\Phi}R^{-2}(e^{-2a\Phi}R^2B_{23})_{,r}=0 \eqn\polb$$
$$\sigma^2 R^2 \lambda^{-2} B_{23}
+\left(\lambda^2 R^2 B_{03}\right)_{,r}
+R^2 \lambda^2 B_{02}
-2Q^2e^{2a\Phi}R^{-2}(N+L)=0 \eqn\polc$$
$$({d \over dr}+(\ln R/\lambda)_{,r})(B_{23}-L-X)-(\ln R)_{,r}
L+\Phi_{,r}\phi=0 \eqn\pold$$
$$B_{23}-T-L+2 \mu^{-2} X=0 \eqn\pole$$
$$(N-L)_{,r}-(\ln \lambda R)_{,r}L+(\ln\lambda/R)_{,r}N-
B_{03}+2\Phi_{,r}\phi=0 \eqn\polf$$
$$X_{,rr}+2(\ln R\lambda)_{,r}X_{,r}
+\mu^2 (2 \lambda^2 R^2)^{-1}(N+L)+
\sigma^2\lambda^{-4}X=0 \eqn\polg$$
$$\eqalign{
2(\ln R)_{,r} N_{,r}+2(\ln\lambda R)_{,r}(B_{23}-L-X)_{,r}
- \mu^2 (\lambda R)^{-2} T
-B_{02}
&\cr
+(\mu^2+2) (\lambda R)^{-2} N
-2 (\ln R)_{,r} (\ln\lambda^2 R)_{,r} L + 2 \Phi_{,r}{}^2 L
&\cr
+2 \sigma^2 \lambda^{-4} (B_{23}-L-X)
-2 a Q^2e^{2 a \Phi} \lambda^{-2} R^{-4} \phi
+2 \Phi_{,r} \phi_{,r}
& = 0}\eqn\polh
$$
$$\eqalign{
R^{-2}(R^2\lambda^2\phi_{,r})_{,r}+\left(\sigma^2\lambda^{-2}-(\mu^2+2)R^{-2}+
2a^2Q^2e^{2a\Phi}R^{-4}\right)\phi+& \cr
\lambda^2\Phi_{,r}\left(N-3L-2X+2 B_{23}\right)_{,r}
-2R^{-2}(R^2\lambda^2\Phi_{,r})_{,r}L+
\lambda^2aB_{02}&=0 \cr}\eqn\poli$$

The two algebraic equations
can be substituted immediately into the other differential equations,
leaving one second order and five first order differential
equations (ignoring the redundant equation \polg\ ) for the six functions
$N$, $L$, $X$, $B_{03}$, $B_{23}$ and $\phi$.

\section{Particular Integral and Reduction of the Polar Equations}

Equations \pola\  through \polh\  reduce to the equations for
the Reissner-Nordstr\"om perturbations given in Chandrasekhar ([\chand ]
page 232)
in the limit $a=0$. The subsequent procedure spelt out in [\chand ] is
somewhat mysterious and
lacks a simple physical interpretation.
It is described as a ``remarkable fact that the system of equations of order
five ... can be reduced to two independent
equations of second order''(p.~233).
This is achieved by defining {\it ad hoc\/} two
nontrivial radius-dependent linear combinations
of the perturbation functions
that can be verified to reduce the system of differential equations.
It was recognized that the origin of this reduction must lie in ``some deeper
fact at the base of the [differential] equations'' (p.~149) and that it is
connected with the existence of a particular solution to the set of equations.
Some very clever work of Xanthapoulos [\xanth ] gives an algorithm which
determines this  particular integral and
thereby enables the construction of the general solution, once the
linear combinations which reduce the equations are known. It
is not at all clear how to generalize
this inspired guesswork to our more general case, so
it was imperative for us to find a rational basis for the procedure.

We will argue that the particular integral is
in fact a manifestation of the fact that the gauge
has not yet been completely set.
This fact can be exploited to find the
particular integral and thereby to reduce the order of the equations,
independent of any prior knowledge of the solution.

Coordinate transformations change the  form of the metric without any
physical
effect. Perturbations which can be undone by a coordinate transformation
therefore automatically
satisfy the equations of motion. One might think that
writing the metric in the form \metric\  eliminates all spurious degrees of
freedom by specifying a unique coordinate system. This is not the case,
however.
There still are coordinate transformations which leave the metric in the form
\metric\ but transform its
individual components. Such transformations are necessarily
independent of $\varphi$, since the gauge choice \metric\ singles out $\varphi$
as the angle corresponding to axial symmetry.

The infinitesimal coordinate transformation
$$x^\mu=\tilde x^\mu+\epsilon \delta x^\mu(\tilde x^\mu) \eqn\tranfo$$
changes the unperturbed metric $g_{\mu\nu}$ to
$$\tilde g_{\mu\nu}=g_{\mu\nu}+
\epsilon {\partial g_{\mu\nu} \over \partial x^\alpha} \delta x^\alpha+
g_{\alpha\beta} {\partial x^\alpha \over \partial \tilde x^\mu}
{\partial x^\beta \over \partial \tilde x^\nu}.\eqn\transmetg$$
For the metric \forme\  we thereby obtain:

$$\delta g_{\mu\nu}=\epsilon \left(\matrix{
\sm{ 2 \lambda^2(\lambda^{-1} \lambda_{,r}\delta r+ \delta t_{,t})}&
\sm{0}&
\sm{{\lambda^2} \delta t_{,r} -  \lambda^{-2} \delta r_{,t}}&
\sm{{\lambda^2} \delta t_{,\theta} - {R^2} \delta \theta_{,t}}\cr
\sm{0}&
\sm{\hskip-32.pt -2 R^2 \sin\theta^2( \cot \theta \delta \theta +
R^{-1} R_{,r} \delta r )\hskip -32.pt}&
\sm{0}&
\sm{0} \cr
\sm{{\lambda^2} \delta t_{,r} -  \lambda^{-2}  \delta r_{,t}}&
\sm{0}&
\sm{ - 2  \lambda^{-2}(\lambda_{,r} \lambda^{-1}\delta r+  \delta r_{,r})}&
\sm{- \lambda^{-2} \delta r_{,\theta}- {R^2} \delta \theta_{,r}}\cr
\sm{{\lambda^2} \delta t_{,\theta} - {R^2} \delta \theta_{,t}}&
\sm{0}&
\sm{-\lambda^{-2} \delta r_{,\theta}- {R^2} \delta \theta_{,r}}&
\sm{\hskip-13.pt -2 R^2 (R^{-1} R_{,r} \delta r+ \delta \theta_{,\theta}})
}\right)\eqn\transmete$$

We must demand that the off-diagonal metric components vanish lest the gauge
choice \metric\ be destroyed. The resulting three differential equations in
$\delta t$, $\delta r$ and $\delta \theta$ are readily integrated
to yield (taking again an $e^{i\sigma t}$ time dependence):
$$\delta t = {R\over \lambda} c(\theta), \qquad
\delta r={\lambda^4 \over i\sigma} \delta t_{,r} \quad {\rm and} \quad
\delta \theta={\lambda^2 \over i \sigma R^2} \delta
t_{,\theta}.\eqn\transsola$$

If we choose $c(\theta)=P_l(\theta)$ and compare \transmete\ with the
definitions \psub\ of
the radial functions, we obtain the following particular integral:
$$\eqalign{
N^{(0)}&=-\sigma^2 R \lambda^{-1} +
\lambda^3 \lambda_{,r} \left(R \lambda^{-1} \right)_{,r} \cr
L^{(0)}&=\left( \lambda^4 \left(R \lambda^{-1} \right)_{,r} \right)_{,r}-
\lambda^3 \lambda_{,r} \left(R\lambda^{-1} \right)_{,r} \cr
T^{(0)}&=\lambda^4 R^{-1} R_{,r}  \left(R \lambda^{-1}\right)_{,r}\cr
X^{(0)}&={\mu^2 \lambda \over 2 R}\cr
B_{23}^{(0)}&=T^{(0)}+L^{(0)}-2 \mu^{-2} X^{(0)}\cr
\phi^{(0)}&=\Phi_{,r} P_l(\theta)^{-1} \delta r =
\Phi_{,r} \lambda^4 \left(R \lambda^{-1}\right)_{,r} }\eqn\parti$$

Given a particular solution to a system of differential equations the
reduction
of the system can be done in a standard and, in principle, straightforward
manner. We will sketch in the following a procedure to obtain three second
order polar equations which keeps the algebra to
a minimum, though even then use of
the symbol manipulation program {\it Mathematica\/} proved
invaluable.

It proves convenient to define a new function $S\equiv B_{23}-L-X$
and write $L$ in terms of $S$. If we now write the perturbation functions in
terms of the particular integral and the functions $n, b, x, s$
and $p$ as
$$\eqalign{
N&=N^{(0)} s +n\cr
B_{23}&=B_{23}^{(0)} s + e^{a \Phi} R^{-1} b\cr
X&=X^{(0)} s +R^{-1} x \cr
S&=S^{(0)} s\cr
\phi&=\phi^{(0)} s + R^{-1} p,}\eqn\sub$$
and substitute them into the equations of motion, all terms in $s$
vanish
and \pold\ becomes an algebraic equation which can be solved
for
$s_{,r}$ easily.
Furthermore, combining \polf\ and \polh\ so as to eliminate $n_{,r}$ gives an
equation which can be solved for $n$.

The polar equations can now finally be reduced to
three second order differential equations in $b$, $x$ and $p$.
Equations \polg\
\footnote*{We use \polg\ here for the first time,
and purely for convenience.  One could also obtain a
much longer looking, though ultimately of course
equivalent, second order equation
in $x$ from \polf\ and \polh .}
and \poli\ are already second order equations in X and $\phi$, respectively
and substitution of \polb\ into \polc\ gives a second order equation
for $B_{23}$.

Thus substitution  of \sub , $s_{,r}$, $n$ and their derivatives into
these second order equations results in a system of equations of the
desired form:

$$\left({d \over dr^*}+\sigma^2\right) {\bf Y}={\bf V}{\bf Y}$$
where ${\bf Y}\equiv (b,x,p)^T$ and {\bf V} is a three by three interaction
matrix. This is just a system of three coupled one-dimensional
wave equations.
The fact that no first order derivatives are present
justifies the definition of $b, x$ and $p$ through \sub . This definition
was  suggested by inspection of the second order differential equations
in $B_{23}$, $X$ and $\phi$, which showed that if
$B_{23}$, $X$ and $\phi$ were written as $e^{a \Phi} R^{-1} \tilde B_{23}$,
$R^{-1} \tilde X$ and $R^{-1} \tilde \phi$, respectively, the equations
contain no first order derivatives with respect to $r^*$. It is, however,
quite remarkable that after substitution of $n$, $s_{,r}$ and their
derivatives  --- which do contain first
order derivatives in these functions --- the final
equations do not contain first
order derivatives.

Unfortunately,
the interaction matrix is too complex to be displayed in explicit
form.  (It will be supplied in electronic form upon request.)
Upon
studying it for
particular values of the parameters $a$, mass and charge, we have
found a most
remarkable surprise: even though all components of the
interaction matrix are complicated
functions of the radius, it has radius-independent
eigenvectors.
This means that it can be brought to a diagonal form,
and that its
eigenvalues have the simple interpretation of being the potentials for three
independent modes! Some plots of these potentials for specific values of the
parameters are shown in Figure~\potfig .
The polar potentials, which we will study more
carefully in the next chapter are of the same form as the axial potentials, in
particular, they go to zero close to the horizon and for large radii, and
they are positive in the intermediate region.

As was the case
for the axial equations, also
for the polar equations
the modes with angular momentum less than two
need to be considered separately. This time we expect to have
two modes for $l=1$, since gravitational waves cannot participate, and only
one mode for $l=0$, because electromagnetic waves are no longer possible
either.
That the gravitational mode disappears can be seen explicitly by realizing that
the parametrization of the metric functions \psub\  is redundant for $l=1$,
since in that case
$-P_1(\theta)=P_1(\theta)_{,\theta \theta}=P_1(\theta)\cot \theta$ so that
variations of $X(r)$ can be absorbed by $T(r)$, and independent on
$X(r)$ for $l=0$, where $X(r)$ gets multiplied by $0=P_0(\theta)_{,\theta}$.
For $l=0$ some of the details of the derivation of equations \pola\ through
\poli\ have to be reviewed to realize that the equations \polb , \polc ,
\pole , \polf\ and \polg\ are in fact satisfied identically because they are
really multiplied by $P_l(\theta)_{,\theta}$. Moreover, there is now a much
larger family of coordinate transformation which don't violate the gauge choice
\metric , namely arbitrary $\delta r$ is allowed which can be used to
set, for example, $T=0$. The four remaining polar equations can then be
reduced to a wave equation for the dilaton field alone.

\section{Magnetically Charged Dilaton Black Holes}

We will now show that the  magnetic case can be
treated in exactly the same manner as the electric case.

As mentioned in chapter 2, magnetically charged solutions
are obtained from the electric solution through the
duality transformation
\forml .
If this transformation is regarded
as a mere change of variables and substituted
into the equations of motion, one finds that the form of the equations of
motions is unaltered, except that unprimed quantities are replaced by
primed
ones. $F'$ can therefore be regarded as an electromagnetic field tensor in its
own right and will correspond to a uniform magnetic field if $F$ corresponded
to a uniform electric field. Conversely, starting with the magnetic solution
and making the change of variables \forml , the primed quantities will
{\it look\/} like the electric solution, but
of course they are still magnetic since
a change of variables has no physical effect.
The perturbation analysis in terms
of the primed quantities is {\it identical\/} to our previous
analysis of the electric case,
and can be taken over without modification. At
the end the unprimed physical quantities are restored by inverting the change
of variables \forml , which has as
its consequences the interchange of electric and
magnetic  perturbations  and change in the sign of the
dilaton perturbation.
For example, $\delta F_{02}$, which transforms to
$e^{-2a\Phi '} F'_{13}$,
will now appear in the axial perturbations.

\section{Remark on stability}

As a by-product of our analysis so far, we can address the question
whether dilaton black holes are stable classically (\ie\
ignoring Hawking
radiation). From the fact that all potentials are positive outside the
outer horizon, one may infer stability of the solution in this region
using the same straightforward argument as
employed
by Chandrasekhar [\chand ] for the special
case of Reissner-Nordstr\"om black holes.
Except for the classic case
$a=0$, $r_-$ is a curvature singularity.  Thus subtleties
concerning instability
of the inner horizon, which occur for classic
Reissner-Nordstr\"om black holes, generically do not arise.

\chapter{Qualitative Features of the Potentials}

In the previous chapter we have found that the equations governing the
perturbations of the metric, electromagnetic, and dilaton fields
can
be reduced to five wave equations for five independent modes.
These modes consist
of various linear combinations of the original
functions parametrizing the perturbations, whose direct physical
meaning is not transparent.
Since the potentials are too unwieldy
to allow a useful description in closed form
and analytical analysis, we will start by discussing the potential of a
simpler, model problem.  We shall consider a
spectator scalar field propagating in the background of a charged dilaton black
hole.  As we shall see, this simple case
seems to display the same main qualitative characteristics as
the other, more  intrinsic potentials.

\section{Spectator Scalar Field}

A massless, uncharged scalar field propagates in curved space according to
the wave equation:
$${1\over \sqrt{-g}} \partial_\mu
\left(\sqrt{-g} g^{\mu\nu} \partial_\nu \eta \right)=0. \eqn\speca$$
This equation reduces to a one-dimensional wave equation in $r^*$
if we make the ansatz
$\eta=e^{i\sigma t} P_{l}(\theta)R^{-1} \tilde \eta$.
The effective potential which appears in this equation is
$V_\eta =R^{-1} R_{,r^*r^*}+l(l+1)\lambda^2 R^{-2}$.  Upon substitution
of the metric functions, this becomes:

$$V_\eta=V_{\eta 1} V_{\eta 2},\eqn\specb$$
where
$$\eqalign{
V_{\eta 1}&=\left( 1 - {r_+\over r} \right)
\left( 1 - {r_-\over r} \right)^ {1 - 3 a^2\over 1 + a^2}\cr
V_{\eta 2}&={1\over r^2} \left(l(l+1)
+{ r_- r + r_+ r \left(1+a^2\right)^2 - \left(2+a^2\right) r_- r_+ \over
\left( 1+a^2 \right)^2 r^2 } - {a^4 r_- \left( 1 - {r_+\over r} \right) \over
 {{{\left( 1 + a^2 \right) }^2} r \left( 1 - {r_-\over r} \right) }} \right)
}\eqn\specc$$

The most
interesting features of this potential arise from the factor $V_{\eta 1}$.
$V_{\eta 2}$ is positive definite and finite
for all radii larger than the radius of the outer horizon $r_+$.
This is
the region we are mainly
interested in, because the region inside the outer horizon
is inaccessible to the outside world.

We are particularly interested in examining what happens as the black hole
becomes extremally charged. For all non-extremal black holes, the potential
vanishes at the horizon by virtue of the factor $V_{\eta 1}$, and is finite
outside the horizon.
Considering even for a moment the extremal limit $r_-=r_+$,
in which $V_{\eta 1}$
becomes $\left(1-{r_+ \over r} \right)^{2-2a^2 \over 1+a^2}$,
it becomes clear that here too, as in the description of the thermal
behavior, three cases must be distinguished: $a<1$, $a=1$ and
$a>1$.
\item{\hbox{$a<1$:}}
The potential is qualitatively the same as for a Reissner-Nordstr\"om
black hole.
For any charge not exceeding the extremal charge,
it is zero at the outer horizon, has a finite maximum and goes back to zero
for large radii.
\item{\hbox{$a=1$:}}
Now we have $V_{\eta 1}={r-r_+ \over r^2 (r-r_-)}$ and the potential is
finite in the strictly extremal limit. Just before becoming extremal, the
potential rises from zero at the horizon $r_+$ to its maximum
value, which it attains at $r-r_+ \sim r_+-r_-$, with an
ever increasing slope. The maximum of the potential
therefore approaches the
horizon as the black hole becomes extremally charged. This means
that in terms of the tortoise $r^*$-coordinate, which is after all
the relevant coordinate for describing the
propagation of waves, the potential becomes
infinitely wide (we may recall that the horizon lies at $r^*=-\infty$
for $a\leq 1$.)
We therefore may say that a finite mass gap develops as the black hole
becomes extremally charged.  More precisely,
excitations with less than a critical frequency are reflected with
certainty.
\item{\hbox{$a>1$:}}
In this case, as we have seen above, the potential diverges at the horizon
in the extremal limit. For non-extremal black holes the potential is, of
course finite outside the horizon, but it has a maximum
whose height grows as
$\left(r_+-r_-\right)^{-2{ a^2 -1 \over 1+a^2}}$.
It would be a little too hasty
to
conclude directly from this diverging
behavior that the transmission coefficient
goes to zero, since
for $a>1$ the horizon lies at a finite value of $r^*$ in the extremal
limit.
Indeed, the potential decreases in width as it increases in height.
An estimate of the WKB-integral $\int \sqrt{V_\eta} dr^*$, however,
yields
a transmission amplitude which behaves
as $e^{-c\ln |r_+-r_-|}$, with $c$ a positive
constant, so that the transmission
probability for any fixed frequency vanishes as the black hole
approaches extremality.
This result
is substantiated by numerical integration of the wave-equation.
So in this precise sense, there is an infinite mass gap for $a>1$.

\section{Axial and Polar potentials}

Although a similar discussion of the qualitative
features of the axial and polar potentials could be given at this point,
it would be
little more than
several repetitions
of the same story we have just told,
in a considerably more cumbersome
form.  Instead we offer the plots  of the axial and polar
potentials in comparison with the spectator potentials displayed in
Figure~\potfig ,
for a variety of parameters. The potentials are seen to exhibit the
same characteristics described in the previous section for the spectator field.
Sampling the potentials
of a few other partial waves confirms that the potentials increase, as
might have been
expected, with increasing angular momenta,
but introduce no obvious qualitative changes.
In particular, since angular momentum enters the equations through the
factor $\mu^2 \lambda^2 R^{-2}$, no infinite centrifugal barrier
develops for $a\leq 1$
as the black hole becomes extremally charged,
even though the area of the two-sphere located at the horizon
tends towards zero (for $a>1$ the potential diverges in all modes anyway).

\chapter{Emission Spectrum}

The rate of emission for a given
bosonic mode is, according to Hawking [\hawkinga ],
equal to
$$
\Gamma_n \left(e^{\sigma/T}-1\right)^{-1} \eqn\spectrum
$$
where $\Gamma_n$ is the absorption coefficient for the  mode $n$ and $T$
the temperature of the black hole.
Given a wave equation with a potential barrier, the absorption coefficient
is calculated in a straightforward manner: Imposing the boundary condition
of incoming waves close enough to the horizon so that the potential is
negligible, the wave equation is integrated outward to a region where the
potential is again negligible and separated into incoming and reflected wave
amplitudes.
The absorption coefficient $\Gamma_n$ is then determined through
$\Gamma_n=1-|{\cal A}_{in}|^2/|{\cal A}_{ref}|^2$.
Figure \absfig\
displays an example
of the dependence of the absorption coefficient on frequency. As expected
the absorption coefficient drops rapidly to zero for frequencies below the
potential barrier. Furthermore we display in Figure \specfig\
the probability of emission of the
spectator field in the $l=2$ mode as obtained from \spectrum\
for various values of parameters on a semilogarithmic plot.
Only the spectator spectrum is shown, but due to the similarities in the
potentials, the spectrum for emission of gravitational, electromagnetic and
dilaton waves look very similar.

It should be noted that compared to a black body,
the spectrum of a black hole is seriously distorted in the low
frequency regime and the total emission rate is reduced by several orders
of magnitudes for all values of the parameters. This is because the
potential barrier, even
in the most favourable cases, is
several times larger then the average available
thermal energy $T$.  This has the important physical consequence, as
discussed in [\thorne ] and
further developed and exploited in [\teit ], that
it makes physical sense to think of the black hole in isolation
as having a physical
{\it atmosphere\/} at the thermal energy $T$ (or, more precisely,
this divided by the value of ${g_{00}}^{1/2}$) which is only in weak
thermal contact with the outside world.

In view of the issues discussed in Chapter 1,
we again are particularly interested in the extremal limit.
Once more, we must consider
three different cases.  Recalling the formula for the black hole
temperature, we find:
$$T= {1 \over 4 \pi r_+} \left(r_+ -r_-\over r_+\right)^{1-a^2\over 1+a^2}$$
\item{\hbox{$a<1$:}}
The temperature goes to zero as the black hole becomes extremal
($r_+ \to r_-$).
The black hole therefore asymptotically approaches the
extremal limit and the radiation switches off when it is reached.
While the grey body factors drastically slow down the approach to extremality,
they do not alter the physical behavior.  The final state of the
black hole is reasonably described as an extended object, similar to
a liquid drop, for $a=0$.  For $0<a<1$  it very plausibly
represents a unique,
non-degenerate ground state, as indicated by its vanishing
entropy.
When an extremal hole hole does swallow
a small amount of mass, its temperature
will become much higher than the energetic distance back to
the ground state, in view of \thermf .
Nevertheless,
the time delay for re-emission plausibly remains long.
Indeed if we use the thermal picture at least seriously enough to
use it to motivate qualitative {\it bounds\/} on the emission, we can argue
as follows.  The rate of emission can be bounded by that of a perfect black
body, where to be generous we take the radius and temperature
to be the values before the emission.
(As we have mentioned before,
both of these
drop drastically with the emission of a single typical quantum.)
Thus
$$
{dM\over dt} ~\lsim~ AT^4 ~\sim~ (r_+ - r_-)^{4-2a^2\over 1+a^2}
{}~\sim~ (\delta M )^{4-2a^2\over 1+a^2}
\eqn\radbound
$$
using \therma , \thermb , and \thermi .
Thus for the time to re-emit the injected mass $\delta M$
we find
$$
\delta t ~\gsim~ (\delta M )^{-3({1-a^2\over 1+a^2})}
\eqn\logwait
$$
which indeed diverges for small $\delta M$ and $a<1$.
Thus the $0<a<1$ holes behave physically
as extended objects, similar to the classic $a=0$
Reissner-Nordstr\"om holes.  In a way it is gratifying that
the massive degeneracy which appears in the classic case,
and seems
somewhat accidental,
is lifted in the generic case.
\item{\hbox{$a=1$:}}
In this case, the temperature
has a finite value as the extreme case is
reached. As we have shown, simultaneously a finite mass gap develops.
This
heavily suppresses the radiation but doesn't quite stop it.
We have found nothing to prevent
continued radiation, and
formation of a
naked singularity.  One may also recall that
the signature \thermf\  for
breakdown of a thermal description did not work for $a=1$.
Thus this case is quite enigmatic.
\item{\hbox{$a>1$:}}
The temperature now becomes formally infinite as the black approaches its
extremal state. At the same time, however, the potentials grow at the
same rate as the average thermal energy. This in itself would not be sufficient
to turn off the radiation, but the radiation has to come to a halt
nonetheless.
We have already argued on general grounds in section 3.3. that the thermal
description necessarily breaks down as the extremal state is approached because
the emission of a quantum with typical thermal energy induces a large
fractional change in the temperature, rendering
the thermal description
at least ambiguous.
In the case at hand one can see what has
to happen instead. The thermal description was derived
by ignoring the back reaction of the radiation on the metric.
Although it is not yet known how to incorporate the back reaction
in a consistent manner, it certainly seems
reasonable to anticipate that a black hole cannot
possibly radiate matter with energies larger than the black hole's mass.
For quanta with energy less than the mass of the black hole, however,
the rate of emission tends to zero by virtue of the development of an
infinite mass gap.  Hence the radiation
slows down and comes to an end at the
extremal limit, despite the infinite temperature.
The infinite temperature however indicates that there will be no time
delay in the scattering of low-energy quanta, again consistent with
an elementary particle interpretation.

\chapter{The Callan-Rubakov Mode}

There is a special feature of the interaction of elementary particle
magnetic monopoles with minimally charged fermions, which is quite
relevant to the circle of ideas discussed above.  We have been much
concerned with the question, whether various particles -- the particles
in the basic, or spectators -- can reach the singular (for
$a\neq 0$) horizon.  It is a famous fact [\rubakov , \callan ]
that minimally charged fermions in the
total angular momentum zero mode are focused right to the core of
particle magnetic monopoles.  The reasons for this are connected
to anomalies and topology [\goldstone , \callwitt ], and are therefore such
as might be expected to apply even to black holes.  Do they?

The Dirac equation in curved space is written most conveniently in terms of
vierbeins and a connection determined by (see \eg
[\birrell ],
p.~85ff.) :
$$\gamma^a \nabla_a \psi =0,\eqn\diraca$$
where
$$\nabla_a = e_a^\mu (\partial_\mu+ieA_\mu+\Gamma_\mu)\eqn\diracb$$
and
$$\Gamma_\mu={1 \over 8} [\gamma^a,\gamma^b] e_a^\nu (\partial_\mu e_{b\nu}-
  \Gamma_{\mu \nu}^\rho e_{b\rho} ).\eqn\diracc $$
For a magnetically charged black hole of minimal charge $1/2e$
the only non-zero component of the electromagnetic field tensor is
$F_{\theta\varphi}={\sin \theta/2e}$.  This may
be obtained from the
vector potential $A_\mu$ whose only non-zero component is
$A_\varphi=-{\cos\theta /2e}$.
The explicit  Dirac equation for a fermion in the background of a minimally
charged magnetic black hole simplifies
by using, as usual, the tortoise coordinate $r^*$
and performing the change of
variable $\xi={\psi / R  \sqrt{\lambda}} $.  The equation is then
$$
\left(\gamma^0 \partial_t+{\gamma^1 \lambda \over R \sin\theta}
\partial_\varphi +
\gamma^2 \partial_{r^*} + {\gamma^3 \lambda \over R} \partial_\theta +
{\cot\theta \lambda \over 2 R} (-i \gamma^1 +  \gamma^3)\right)
\xi=0.
\eqn\diracd$$
In the chiral representation of the $\gamma$-matrices,
$\gamma^0=\left(\matrix{0&-1\cr-1&0}\right)$ and
$\gamma^i=\left(\matrix{0&\sigma^i\cr-\sigma^i&0}\right)$,
the equation
for massless fermions
separates immediately into
two sets of equations for two-spinors.  If for convenience
we identify
$(\sigma^1,\sigma^2,\sigma^3)$ with the standard Pauli-matrices
$(\sigma_y,\sigma_z,\sigma_x)$ and write the two-spinor as
$\left(u \atop v \right)$, these equations become:
$$\eqalign{
\left(\pm \partial _t+\partial_{r^*} \right) u_\pm + {\lambda \over R}
\left( {i \over \sin \theta} \partial_\varphi+\partial_\theta +\cot \theta
\right) v_\pm&=0\cr
\left(\pm \partial _t-\partial_{r^*} \right) v_\pm + {\lambda \over R}
\left(- {i \over \sin \theta} \partial_\varphi+\partial_\theta \right)
u_\pm&=0,}\eqn\dirace$$
where $\pm$ distinguishes fermions of opposite chirality.
Acting with $${1\over \sin^2\theta} \left( i \partial_\varphi
+ \sin \theta
\partial_\theta \right) \sin \theta$$
from the left on the second equation
and substituting into the  first equation,
when
the angular dependence of $u$ is taken to be $P_l(\theta)$,
we arrive at the following equation:
$$\left(\pm \partial_t -\partial_{r^*}\right) {R\over \lambda}
\left(\pm \partial_t +\partial_{r^*}\right) u_\pm+{\lambda \over R} l(l+1)
u_\pm=0 ~. \eqn\diracf$$
We see that
fermions are
in general subject to a frequency dependent potential.  This occurs also
for
Kerr black holes, and is a sign of the underlying
time-reversal asymmetry of the problem.  (Magnetic charge is
T odd.)

More significantly, we see that the
characteristic behavior of the Callan-Rubakov mode survives the transition
from standard to black hole monopoles.
For zero angular momentum no potential at all is
present -- regardless of the black hole parameters!
The possibility of
modes without any potential is particularly significant
for dilaton black holes, especially those with
$a>1$.  This mode is the only one that can penetrate to the
core of an extremally charged black hole in this regime.
In view of the
formally infinite temperature of the black hole,  it would seem
at the classical level that catastrophic radiation would ensue.
We believe that a proper quantization will identify a stable
ground
state and a discrete spectrum of dyonic excitations
around it, similar to what occurs for particle monopoles in
flat space -- but this is a problem for the
future.

\chapter{Final Comments}

Although falling well short of a derivation,
we believe that the calculations and arguments
given above
provide substantial evidence for the
consistency of the physical picture
outlined in the introductory section.
An important qualitative
conclusion is that extremal black holes
with the same quantum numbers can, for
different field contents of the world, represent
rather different physical entities.  The extremal
charged dilaton black holes appear to be extended spherical
objects for $a<1$, and elementary point objects for $a>1$.
Unfortunately the case suggested by superstring theory,
$a=1$, is enigmatic.  However, we can
hardly refrain from observing
that
a string is the intermediate case between a spherical membrane and
a point!

\ack{We are very grateful to Sandip Trivedi and to Edward Witten
for several valuable comments and suggestions, and to Gerald
Gilbert for discussion of his calculations.}

\endpage

\refout

\endpage

\figout

\end

\end